# A Differential Evolution Algorithm with Neighborhood Mutation for DOA Estimation

Bo Zhou[a,b,c], Kaijie Xu[a,b,c,d], Yinghui Quan[a,b,c,d] and Mengdao Xing[a,b,c,d]

[a]*School of Information Mechanics and Sensing Engineering, Xi'an 710071, China.*

[b]*School of Electronic Engineering, Xidian University, Xi'an 710071, China.*

[c]*Key Laboratory of Collaborative Intelligence Systems, Ministry of Education, Xi'an 710071, China*

[d]*Hangzhou Institute of Technology, Xidian University, Xi'an 710071, China.*

**ABSTRACT**

Two-dimensional (2D) Multiple Signal Classification algorithm is a powerful technique for high-resolution Direction Of Arrival (DOA) estimation in array signal processing. However, the exhaustive search over the 2D angular domain leads to high computational cost, limiting its applicability in real-time scenarios. In this study, we reformulate the peak-finding process as a multimodal optimization problem, and propose a Differential Evolution algorithm with Neighborhood Mutation (DE-NM) to efficiently locate multiple spectral peaks without requiring dense grid sampling. Simulation results demonstrate that the developed scheme achieves comparable estimation accuracy to the traditional grid search, while significantly reducing computation time. This strategy presents a promising solution for real-time, high-resolution DOA estimation in practical applications. The implementation code is available at https://github.com/zzb-nice/DOA_multimodel_optimize.

## 1. Introduction

Array signal processing has emerged as a critical technique in modern signal analysis, enabling precise parameter estimation, source localization, and interference suppression across a wide range of engineering fields [1]–[4]. While traditional Direction Of Arrival (DOA) estimation methods typically focus on a single angular dimension, many practical scenarios such as three-dimensional target localization require joint estimation of spatial parameters in both azimuth and elevation. This has led to an increased interest in two-dimensional (2D) DOA estimation techniques.

In practical applications, two-dimensional Multiple Signal Classification (2D-MUSIC) algorithm [5] and 2D estimation of signal parameters via rotational invariant techniques (2D-ESPRIT) algorithm [6] are widely employed for 2D DOA estimation. Among these, 2D-MUSIC is generally regarded as more robust and accurate than 2D-ESPRIT [7][8]. This superior performance is largely attributed to its exhaustive spectral search mechanism, which, while computationally intensive, enables precise identification of multiple sources across both azimuth and elevation dimensions.

Due to its high resolution and robustness, the 2D-MUSIC algorithm has been extensively utilized in various signal processing systems. In orthogonal frequency division multiplexing (OFDM)-based sensing systems, the 2D-MUSIC algorithm serves as an effective approach for estimating range-Doppler [9] and range-angle [7][10] parameters. In impulse radio ultra-wideband (IR-UWB) positioning systems, 2D-MUSIC algorithm is applied to the joint estimation of time-of-arrival (TOA) and DOA [8][11][12]. Similarly, the 2D-MUSIC algorithm is also frequently employed in frequency-modulated continuous-wave (FMCW) signal processing[13]–[15].

However, the conventional 2D-MUSIC algorithm involves an exhaustive grid search over the entire parameter space, resulting in high computational complexity. This characteristic significantly restricts its applicability in real-time or resource-limited systems. Consequently, the development of low-complexity 2D-MUSIC algorithms is essential to enable practical implementation in modern signal processing applications.

A variety of methods have been introduced into the 2D-MUSIC algorithm to reduce its computational complexity. Dimensionality-reduced MUSIC approaches [16][17] can transform the high-dimensional spectral peak search in 2D-MUSIC into a one-dimensional search through certain processing techniques, thereby significantly decreasing the computational burden of the model. However, such methods generally rely on specific array structures, rendering them unsuitable for handling existing sparse arrays or arbitrarily shaped arrays. In contrast, array transformation methods [18][19] aim to convert arbitrarily shaped two-dimensional arrays into two uniform linear arrays (ULAs). These transformed arrays allow the application of root-finding techniques, thereby circumventing the exhaustive spectral search. Nonetheless, this transformation process may introduce model mismatch errors, which could degrade the accuracy of the DOA estimation.

Through approaches such as gradient descent [20] or swarm intelligence [21]–[23], certain optimization algorithms are capable of iteratively computing the extrema of an objective function. These methods have seen widespread adoption in engineering applications and have been extensively validated. If the 2D-MUSIC spectrum peak search problem is reformulated as an optimization problem, it becomes possible to leverage such algorithms to simultaneously locate multiple spectral peaks. This approach has the potential to reduce the computational complexity associated with traditional spectrum peak search methods.

Traditional optimization algorithms typically aim to locate a single global optimum of an objective function. These methods are generally categorized into two major classes: gradient-based algorithms and population-based algorithms.

Gradient-based algorithms, such as gradient descent, Newton's method, and quasi-Newton methods (e.g., BFGS), rely on analytical properties of the objective function, including smoothness and differentiability, to converge to a local or global optimum. These methods are computationally efficient and exhibit fast convergence when operating near an optimum. Nevertheless, they are susceptible to being trapped in local minima when applied to non-convex or multi-modal functions.

To overcome the limitations of gradient-based optimization algorithms, a broad class of population-based optimization methods has been developed. Representative algorithms in this category include Genetic Algorithms (GA), Differential Evolution (DE), Particle Swarm Optimization (PSO), and Sparrow Search Algorithm (SSA), which are inspired by natural selection mechanisms and collective behaviors observed in biological systems. Unlike gradient-based methods, these algorithms are capable of efficiently exploring irregular search spaces.

The 2D-MUSIC spectrum peak search problem requires the simultaneous identification of multiple prominent spectral peaks, which naturally formulates the problem as a multimodal optimization task. Multimodal optimization algorithms are designed to identify several high quality global or local solutions of the objective function. Compared with traditional optimization, multimodal optimization presents substantially greater challenges. It is required not only to maintain high precision in peak localization, but also to ensure that all meaningful local optima are effectively captured, as failure implies the loss of a true peak corresponding to an actual signal source in the 2D MUSIC algorithm.

To achieve this, a variety of strategies have been proposed to enable traditional optimization algorithms to maintain diversity and avoid premature convergence. Notable techniques include fitness sharing, crowding, speciation, clearing, and memory-based archiving, all of which are designed to preserve the stability of the existing local convergence while encouraging extensive exploration of the overall search landscape.

Building on these strategies, several multimodal variants of classical optimization algorithms have been developed, such as clearing GA [24][25], sharing GA [26], crowding-based Differential Evolution (CDE) [27]–[29], and Species-based PSO (SPSO) [30]. Although these methods have been primarily applied to benchmark functions and simple multimodal tasks, they have laid a solid foundation for tackling more complex and real-world multimodal optimization problems.

In recent studies, only a few methods have explicitly reformulated the 2D-MUSIC spectral peak search problem as an optimization task, aiming to reduce the computational burden associated with exhaustive grid search. In 2025, Hu et al. [7] developed a Dung Beetle Optimization (DBO) algorithm to accelerate the 2D-MUSIC procedure. However, this population-based iterative method was designed to identify only a single spectral peak, limiting its applicability in multi-source scenarios. Earlier, in 2019, Zhu et al. [31] introduced a Dandelion PSO (DPSO) algorithm for the same task. Their method incorporated the Basic Sequential Algorithmic Scheme (BSAS) to cluster particles into subpopulations and adopted a dandelion-inspired seed dispersal mechanism to preserve high-quality solutions during the search process.

However, the existing methods rarely consider or evaluate the success rate of spectral peak detection algorithms. In fact, even well-established global optimization methods may exhibit a non-negligible failure rate in reliably locating the global optimum. This challenge becomes more pronounced in the context of multi-peak spectral estimation for 2D-MUSIC, where the search space is more complex and the optimization task becomes significantly more difficult. Such scenarios require optimization algorithms to demonstrate greater robustness, strong global exploration capability, and fast convergence, which is a highly demanding engineering task.

In this study, a Differential Evolution algorithm with Neighborhood Mutation (DE-NM) [32] is developed to address the challenges associated with spectral peak searching in 2D-MUSIC. DE is a widely used population-based optimization method, known for its conceptual simplicity and strong performance across various application domains. To extend DE to multimodal optimization problems, a neighborhood-based mutation strategy is introduced, which restricts mutation operations to a set of distance-based neighboring solutions. This restriction guides the population toward diverse regions of the solution space during iteration, thereby enabling the algorithm to effectively identify multiple local optima.

Then, the Density-Based Spatial Clustering of Applications with Noise (DBSCAN) algorithm [33] is employed as a post-processing step to aggregate the final set of solutions into spatial clusters. Each spatial cluster is regarded as corresponding to a spectral peak in the 2D-MUSIC spectrum, with the final estimate derived from the solution having the highest fitness within the cluster. To the best of our knowledge, existing population-based approaches for 2D-MUSIC have not addressed how to extract peak estimates from the final population. This work is the first to introduce a clustering-based strategy for spectral peak extraction, rather than directly selecting the sample closest to the true peak in simulation. This strategy provides a truly practical and deployable algorithm, as opposed to prior methods that limited to simulation scenarios, thereby offering substantial real-world significance.

Experimental results demonstrate that the proposed DE-NM method consistently outperforms conventional multimodal optimization methods in terms of robustness and convergence speed, achieving strong overall performance.

The rest of this paper is organized as follows. The signal model is presented in Section 2. Section 3 describe the complete workflow of the proposed DE-NM algorithm. Extensive simulations are given in Section 4 and conclusions are summarized in Section 5.

## 2. PROBLEM FORMULATION

Consider a Uniform Circular Array (UCA) or Uniform Rectangular Array (URA) consisting of $M$ omnidirectional antenna elements, receiving $L$ far-field narrowband source signals. The source signal vector is defined as $s(t) = [s_1(t), s_2(t), \ldots, s_L(t)]^T$, where $s_l(t)$ corresponds to the signal from the $l$-th source. And the corresponding DOAs are denoted as $(\boldsymbol{\theta}, \boldsymbol{\varphi}) = [(\theta_1, \varphi_1), (\theta_2, \varphi_2), \ldots, (\theta_L, \varphi_L)]^T$, where $\theta_l$ and $\varphi_l$ represent the azimuth and elevation angles of the $l$-th source, respectively. The array received signal model can be expressed in the following manner:

$$X(t) = A(\boldsymbol{\theta}, \boldsymbol{\varphi})s(t) + n(t), t = 1, \ldots, T \quad (1)$$

where $A(\boldsymbol{\theta}, \boldsymbol{\varphi}) = [a(\theta_1, \varphi_1), a(\theta_2, \varphi_2), \ldots, a(\theta_L, \varphi_L)]$ denotes the array manifold matrix, composed of steering vectors corresponding to each DOA. The noise vector $n(t)$ is assumed to have a zero-mean and is independent of the observed signal. For a UCA, $a(\theta, \varphi)$ can be represented in the following form:

$$a(\theta_l, \varphi_l) = \left[e^{-j\frac{2\pi}{\lambda}cos(\phi_1 - \theta_l)sin(\varphi_l)}, \ldots, e^{-j\frac{2\pi}{\lambda}cos(\phi_M - \theta_l)sin(\varphi_l)}\right]^T \quad (2)$$

where $\phi_m = \frac{2\pi m}{M}$ represents the azimuthal angle of the $m$-th element in the UCA, with $M$ representing the total number of sensors. To extract the spatial information from the received data, the sample covariance matrix $R \in \mathbb{C}^{M \times M}$ is constructed as:

$$R = E[X(t)X^H(t)] \approx \frac{1}{T}\sum_{t=1}^{T} X(t)X^H(t), \quad (3)$$

where $(\cdot)^H$ denotes the Hermitian transpose, $E[\cdot]$ is the expectation operator, and $T$ is the number of snapshots used in the estimation. The covariance matrix captures both signal and noise characteristics and serves as the fundamental input for subspace-based DOA estimation algorithms such as MUSIC. To facilitate the separation of the signal and noise components from the covariance matrix $R$, subspace-based methods typically perform an eigenvalue decomposition of $R$ as follows:

$$R = U_s \Lambda_s U_s^H + U_n \Lambda_n U_n^H \quad (4)$$

where $U_s$ and $U_n$ denote the orthonormal eigenvector bases corresponding to the signal and noise subspaces, respectively. The diagonal matrices $\Lambda_s$ and $\Lambda_n$ contain the associated eigenvalues.

## 3. DE-NM ALGORITHM

The proposed algorithm is designed to achieve robust DOA estimation by combining subspace theory, global-local hybrid optimization, and density-based clustering techniques. In this section we present a detailed description of the DE-NM algorithm. The motivation and design rationale of the proposed DE-NM algorithm are first discussed in Subsection A.

Subsequently, we present an overview of the algorithm and provide a systematic description of each fundamental component. As shown in Figure 1, the algorithm is structured into four principal phases, with the main components detailed in Subsections B-D.

### A. Motivation Behind the Proposed Method

Motivated by the practical demand in engineering applications for DOA estimation algorithms that are both accurate and computationally efficient, this work focuses on enhancing the performance of the 2D-MUSIC algorithm. Although 2D-MUSIC offers high-resolution DOA estimation, its reliance on exhaustive grid search leads to substantial computational overhead.

To address this limitation, we reformulate the spectral peak search as a multimodal optimization problem and introduce a tailored algorithmic framework based on differential evolution. This approach aims to reduce computational cost while maintaining high estimation accuracy, thereby improving the algorithm's suitability for real-time and resource-constrained environments.

Moreover, existing population-based optimization approaches applied to 2D-MUSIC seldom discuss how to extract DOAs from the final population distribution, which is a critical step for practical implementation. Without an effective postprocessing strategy, the algorithm may fail to produce reliable estimation results, even when multiple peaks are successfully identified via multimodal optimization.

In our approach, the DBSCAN algorithm is employed in a novel way as a postprocessing method to extract DOAs from the solution distribution. The resulting estimates are compared with two baseline methods: (1) **k-localmax** and (2) **k-means++**. Experimental results show that the proposed postprocessing method achieves improved robustness during the DOAs extraction stage.

**B. Noise Subspace Projection Matrix Calculation**

The foundation of this step lies in subspace theory, especially the 2D-MUSIC algorithm. After decomposing the sample covariance matrix $R$ into the signal and noise subspaces, the 2D-MUSIC algorithm constructs a pseudo-spectrum to evaluate the orthogonality between a candidate steering vector and the noise subspace. The spectrum is given by:

$$P_{\text{MUSIC}}(\theta, \varphi) = \frac{1}{a^H(\theta, \varphi) \, U_n U_n^H \, a(\theta, \varphi)} \quad (5)$$

Local maxima in this spectrum correspond to directions where the steering vector is nearly orthogonal to the noise subspace, indicating potential signal sources.

To avoid repeated matrix multiplications in practice, the noise subspace projection matrix $G_n = U_n U_n^H$ is typically pre-computed. This allows the MUSIC spectrum to be reformulated as:

$$P_{\text{MUSIC}}(\theta, \varphi) = \frac{1}{a^H(\theta, \varphi) \, G_n \, a(\theta, \varphi)} \quad (6)$$

However, the exhaustive grid search over a high-resolution two-dimensional parameter space is computationally intensive. Each evaluation of the MUSIC spectrum requires approximately $M^2 + M$ floating-point operations. For a grid of $N_\varphi \times N_\theta$ points, the total computational cost becomes significant, especially as the resolution increases.

In the developed scheme, the 2D-MUSIC spectrum is regarded as an objective function, and a multimodal optimization strategy is employed to efficiently identify multiple spectral peaks, with the aim of reducing the computational overhead induced by exhaustive peak search.

**C. Differential Evolution with neighborhood-based mutation strategy**

The DE algorithm is a simple yet robust global optimization method that has been successfully applied in a wide range of fields [29]. Unlike traditional evolutionary algorithms, DE updates individuals by employing the differences between randomly selected pairs of vectors from the population. This strategy enables the algorithm to self-adaptively adjust the search step size based on the distribution of the current population, thereby enhancing global exploration and reducing the risk of premature convergence.

A simplified pseudocode of the DE algorithm is presented in Table 1 to illustrate its core procedures. The performance of the DE algorithm is influenced by four primary parameters: the population size $P$, scaling factor $F$, crossover rate $CR$, and maximum number of iterations $Max\_iter$. These parameters collectively control the balance between exploration and exploitation, convergence speed, and computational efficiency of the algorithm. For instance, the scaling factor $F$ controls the mutation step size through $v_i = x_{r1} + F \cdot (x_{r2} - x_{r3})$, directly affecting the algorithm's ability to explore the search space. Similarly, $CR$ influences the diversity of offspring by determining how components are exchanged between individuals.

However, the mutation method is suitable for solving single global optimum optimization problems but is inappropriate for solving multimodal optimization. To address the challenge of multimodal optimization, our proposed DE-NM algorithm enhances the standard DE by introducing a neighborhood mutation mechanism. The pseudocode of the proposed method is presented in Table 2.

In the proposed method, a local search is applied around each individual by finding $m$ nearest neighbors to form a local subpopulation. With a neighborhood-based mutation strategy confined to each individual's local subpopulation, particles explore distinct regions of the search space independently. This mechanism helps refine solutions in complex landscapes while preserving diversity, thereby enabling the algorithm to converge toward multiple local optima in the 2D-MUSIC spectrum.

After generating the trial vector through neighborhood mutation and crossover, the selection step follows the standard DE rule: the offspring replaces the parent if it yields a better fitness value.

**D. DBSCAN-based postprocessing**

The preceding DE-NM algorithm has effectively guided the population to converge around multiple local maxima in the search space. However, to obtain a final estimate of $L$ DOAs, a postprocessing step is required to extract representative solutions from the evolved distribution. Since directly selecting individuals from the final population may result in redundancy and sensitivity to noise, a natural solution is to introduce a clustering-based strategy that groups the population into meaningful regions corresponding to potential signal sources.

In the proposed method, the density-based clustering algorithm DBSCAN is adopted to group individuals based on spatial proximity, resulting in $K$ distinct clusters that correspond to dense regions in the solution space. For each detected cluster, the individual with the highest fitness is selected to serve as the representative solution, and the top $L$ representatives will be retained as the final DOA estimates.

Unlike centroid-based methods, the DBSCAN algorithm does not require prior knowledge of the number of clusters and is inherently robust to noise and outliers. Experimental results demonstrate that the proposed method achieves more accurate and stable DOA estimation compared with conventional methods **k-localmax** and clustering-based strategy **k-means++**, exhibiting the best overall robustness and performance under varying noise and source conditions.

## 4. SIMULATIONS AND ANALYSES

In this section, extensive experiments are conducted to demonstrate the superior performance of the proposed method in DOA estimation. A detailed comparison of the DOA estimation performance between the proposed method and existing approaches is provided in Subsection A. In addition, computational complexity analysis is provided to validate the runtime efficiency of the proposed method in Subsection B. Overall, the resulting algorithm offers an effective trade-off between accuracy and computational cost, making it a reliable solution for practical applications.

To demonstrate the effectiveness of the proposed model, a variety of representative algorithms were selected for comparative analysis. For comparative analysis, several well-established multimodal extensions of the classical DE algorithm were implemented. Specifically, Species-based DE (SDE), Deterministic Crowding DE (DC-DE), and Fitness Sharing DE (Sharing-DE) were implemented by incorporating speciation, deterministic crowding, and fitness sharing mechanisms, respectively. Furthermore, the DPSO algorithm [31] and the MUSIC-AP algorithm [34][35] were also implemented for comparative analysis. The DPSO is recognized as a well-established multimodal optimization algorithm and represents one of the most recent approaches applied to the 2D-MUSIC spectrum. In contrast, MUSIC-AP employs a greedy alternating optimization strategy by iteratively updating the parameters $\theta$ and $\varphi$, thereby guiding the algorithm toward the optimal solution.

All algorithms were individually tuned on the 2D-MUSIC spectrum to ensure a fair balance between accuracy and computational efficiency. Moreover, since the original MUSIC-AP algorithm exhibits limited performance in multi-target scenarios, we extend it by introducing multiple initialization points, resulting in behavior similar to that of a population-based approach.

### A. Computational Complexity Analysis

To assess the computational efficiency of the proposed method, a detailed analysis of its computational complexity is provided.

**2D-MUSIC algorithm:** The computational complexity of the conventional 2D-MUSIC algorithm is primarily dominated by two stages: eigen-decomposition of the covariance matrix and exhaustive spectral search over a two-dimensional parameter space. By the application of Fast Subspace Decomposition (FSD) technique [36], the computational complexity of the subspace decomposition step can be reduced to $M^2(L+2)$ floating-point operations (FLOPs).

In addition, let $J = N_\theta \times N_\varphi$ stand for the number of grid points in the parameter space, where $N_\theta$ and $N_\varphi$ are the numbers of angular samples in azimuth and elevation, respectively. Since each spectral evaluation requires $(M+1)(M-L)$ FLOPs, the total computational cost of the spectral search step becomes $J(M+1)(M-L)$ FLOPs. Therefore, the overall computational complexity of the 2D-MUSIC algorithm can be expressed as:

$$C_{MUSIC} = M^2(L+2) + J(M+1)(M-L) \; FLOPs \quad (7)$$

In practical scenarios, the number of grid points $J$ is typically much larger than both the number of sensors $M$ and the number of sources $L$, i.e. $J \gg M > L$. Therefore, the computational burden of the spectral search stage becomes the dominant factor in the overall complexity of the 2D-MUSIC algorithm.

**Population-based algorithm:** To address the high computational cost of the spectral search stage, population-based optimization algorithm is adopted to replace the exhaustive grid search used in conventional 2D-MUSIC. The population size $N_R$ and the maximum number of iterations $Max\_iter$ are two key parameters that significantly influence the performance and computational cost of population-based optimization algorithms.

In each iteration, the optimization algorithm evaluates the fitness of all individuals in the population, which is equivalent to computing the 2D-MUSIC spectrum in the parameter space and requires $(M+1)(M-L)$ FLOPs. Therefore, the total spectral evaluation cost of the population-based algorithm is $N_R \times Max\_iter \times (M+1)(M-L)$ FLOPs.

Moreover, as most population-based algorithms rely on the spatial relationships among individuals to guide the optimization process, computing pairwise distances is often necessary in each iteration, which introduces additional computational overhead. Each pairwise distance computation involves 2 FLOPs. Since there are $\frac{N_R(N_R-1)}{2}$ unique pairs in a population of size, the total cost of distance calculations per iteration is $N_R(N_R - 1)$ FLOPs. Therefore, the overall computational complexity of the population-based algorithm can be expressed as:

$$C_{population} = M^2(L+2) + Max\_iter \times N_R \times \\ ((M+1)(M-L) + (N_R - 1)) \; FLOPs \quad (8)$$

With the values of $Max\_iter$ and $N_R$ properly tuned according to inherent characteristics of the population-based algorithm, it is possible to balance detection accuracy and computational efficiency.

The primary motivation of the proposed method is to improve computational efficiency without compromising the estimation performance. Based on this principle, the algorithm parameters are configured as shown in Step 2 of Figure 1. Under this configuration, the proposed model consistently achieves near-perfect accuracy in spectral peak localization, with a statistical success rate approaching 100%. Notably, its overall statistical performance in DOA estimation even surpasses that of the exhaustive spectral search method.

Moreover, to quantitatively assess the computational efficiency under different array configurations, a detailed comparison of the computational complexity between the conventional 2D-MUSIC algorithm and the population-based approach is provided in Table 3, with results expressed in terms of MFLOPs. The results clearly indicate that, even when achieving near-perfect spectral peak localization accuracy, the proposed method substantially reduces computational overhead, particularly in scenarios involving large-scale arrays and multiple incident sources.

Furthermore, in practical scenarios where faster computation is required, the proposed algorithm can be further accelerated by appropriately reducing the population size $N_R$ and the number of iterations $Max\_iter$. This allows for a further reduction in computational cost while ensuring that the estimation accuracy meets the required performance criteria, thereby achieving a balanced trade-off between complexity and precision. A more detailed analysis of this trade-off will be presented in the following part of this work.

### B. Comparative Analysis of DOA Estimation Performance

A series of simulation experiments were conducted to validate the DOA estimation performance of the proposed method. First, a standard DOA estimation scenario is considered, where a UCA with 12 elements receives three incident signals from directions $\theta_{true} = [30.42°, 120.27°, 240.51°]$ and $\varphi_{true} = [60.39°, 29.42°, 45.55°]$.

**Visual Performance Demonstration:** To provide an intuitive comparison of algorithmic performance, the distributions of all population-based methods under SNR = –5 dB and SNR = 5 dB are visualized in Figure 2 and Figure 3, respectively.

All algorithms are initialized with the same population configuration to ensure a fair comparison. It can be observed that the final population generated by the DE-NM algorithm is tightly concentrated around the three true peak locations, demonstrating the best overall performance among the compared methods. The SDE algorithm also shows convergence around the peak regions.

In contrast, the solution distributions of DC-DE, Sharing-DE, and DPSO are more scattered, making peak extraction more challenging. While the MUSIC-AP algorithm is capable of converging to the true peak locations, it is also prone to convergence at boundary-local optima, which may degrade overall estimation accuracy.

As shown in the Figure 2(e) and Figure 3(c), Sharing-DE and DC-DE algorithm fails to identify one of the true peaks, highlighting a potential limitation of multimodal optimization algorithms in reliably detecting all target modes. This observation indicates the potential risk of peak omission in multimodal optimization, highlighting the necessity of further statistical evaluation of algorithm performance.

**MAE vs. SNR:** To provide a comprehensive statistical evaluation, The Mean Absolute Error (MAE) of the azimuth and elevation angles under varying SNR and numbers of snapshots is presented, as illustrated in Figure 4. The results are averaged over 1,000 independent Monte Carlo trials for each SNR level.

As expected, the MAE exhibits a decreasing trend with increasing SNR. This improvement is mainly attributed to the enhanced accuracy and sharpness of the 2D-MUSIC spectrum at high SNRs, which can be clearly observed from the comparison between Figure 2 and Figure 3. In particular, the proposed DE-NM algorithm shows strong robustness across varying noise levels, consistently guiding the population to converge around all true DOA peaks with minimal dispersion. Compared with other population-based methods, DE-NM achieves tighter clustering and fewer false responses, demonstrating superior accuracy and stability in both low- and high-SNR regimes.

Although SDE and MUSIC-AP exhibit performance comparable to the proposed method under high-SNR conditions, their estimation accuracy degrades considerably in low-SNR environments. DC-DE shows greater robustness across different noise levels; however, its overall performance remains slightly inferior to that of the proposed algorithm. Notably, DE-NM algorithm yields better statistical performance in terms of MAE than the conventional 2D-MUSIC algorithm, which relies on an exhaustive grid search over the parameter space. This indicates that the proposed method can not only accelerates computation but also enhances estimation precision.

To further demonstrate the effectiveness of the proposed algorithm, the cumulative distribution functions (CDF) of the absolute estimation errors for both azimuth $\theta$ and elevation $\varphi$ angles are presented under two representative scenarios: SNR = –5 dB and SNR = 5 dB. As illustrated in Figure 5, the proposed method yields steeper CDF curves, reflecting a higher concentration of low-error estimates and thereby demonstrating superior statistical robustness in DOA estimation.

### C. Justification of Algorithm Design and Parameter Selection

In this subsection, we first conduct a comparative experiment to validate the design of the proposed method. To demonstrate the effectiveness of the DBSCAN-based solution extraction strategy employed in the proposed method, we compare its performance with two alternatives: (1) *k-localmax* and (2) *k-means++*.

***k-localmax*** is an extension of the traditional peak detection strategy used in grid-based spectral search, in which a point is designated as a local maximum if its objective value is greater than that of all adjacent points within a predefined neighborhood on the search grid. However, population-based optimization typically yields irregular solution distributions. To adapt to this setting, we define the neighborhood of each point as its $k$ nearest neighbors in Euclidean space. A point is considered a local maximum if its fitness exceeds that of all its neighbors. Upon identification of all local maxima, the top-$L$ local maxima with the highest fitness values are selected as the final DOA estimates.

In addition, ***k-means++*** serves as a clustering-based peak extraction strategy similar to the proposed DBSCAN-based method. It deterministically partitions the population into $L$ clusters, from which the individual with the highest fitness in each cluster is selected as the corresponding DOA estimation value.

Although the two aforementioned methods constitute reasonable solution strategies, the comparative evaluation in Figure 6 reveals that the proposed DBSCAN-based approach offers significantly enhanced robustness to noise and achieves superior statistical accuracy in DOA estimation.

Moreover, as shown in Figure 7, the DBSCAN algorithm identifies certain isolated points as noise. This property enhances its robustness by effectively filtering out scattered outliers that do not cluster around true spectral peaks, thereby reducing the impact of noise and improving the reliability of the final solution extraction.

To further demonstrate the rationale behind our parameter selection strategy, Figure 7 illustrates how the DOA estimation accuracy and computational complexity of the proposed method vary with different population sizes $N_R$. For each population size, the associated algorithmic parameters were coarsely tuned to ensure reasonable performance.

Each plotted marker corresponds to a specific configuration, where each marker size reflects the corresponding computational cost. The label accompanying each point reports the computational complexity as a percentage relative to that of the conventional 2D-MUSIC algorithm. For reference, the red horizontal line indicates the statistical performance of the exhaustive spectral search in terms of MAE.

From Figure 8, it is evident that using a population size of 256 allows the proposed method to achieve superior accuracy compared with the standard 2D-MUSIC algorithm. Under this configuration, the algorithm successfully identifies nearly all DOA peaks with a success rate close to 100%.

Additionally, the MAE performance exhibits a difference between azimuth $\theta$ and elevation $\varphi$. Specifically, the proposed method surpasses the accuracy of 2D-MUSIC in elevation angle estimation at a population size of 192, whereas in the case of the azimuth angle, a larger population size of 256 is required to achieve comparable superiority. This can be attributed to the uniform grid resolution of 1° used by 2D-MUSIC for both dimensions. Since the $\theta$ range typically spans a wider domain than $\varphi$, this leads to a denser grid and higher resolution for $\theta$, resulting in better performance for azimuth estimation in the conventional approach.

## 5. Conclusion

In this paper, a novel DE-NM algorithm was proposed for efficient and accurate 2D DOA estimation. By integrating a neighborhood-based mutation strategy into the differential evolution framework, the algorithm is capable of simultaneously identifying multiple spectral peaks, making it well-suited for multimodal optimization scenarios. Furthermore, a density-based DBSCAN clustering method was innovatively introduced to extract representative DOA estimates from the evolved population, effectively addressing the challenges posed by noisy or irregular distributions. Extensive simulations demonstrate that the proposed method achieves superior accuracy, robustness, and computational efficiency compared with conventional techniques, including those with an exhaustive spectral search. Additionally, a flexible parameter selection strategy enables a tunable trade-off between estimation precision and computational complexity, making the algorithm practical for a wide range of real-world applications.

**CRediT authorship contribution statement**

**Bo Zhou:** Writing – original draft, Data curation. **Kaijie Xu:** Writing – review & editing, Supervision, Methodology. **Yinghui Quan:** Co-supervision. **Mengdao Xing:** Supervision.

**Declaration of competing interest**

The authors declare that they have no known competing financial interests or personal relationships that could have appeared to influence the work reported in this paper

**Data availability**

Data will be made available on request.

**Acknowledgement**

The authors would like to thank the anonymous referees for their valuable comments and suggestions. This research was supported by the National Natural Science Foundation of China

(Nos. 62101400, 72101075, 72171069 and 92367206), and in part by the Shaanxi Fundamental Science Research Project for Mathematics and Physics under Grant 22JSQ032.

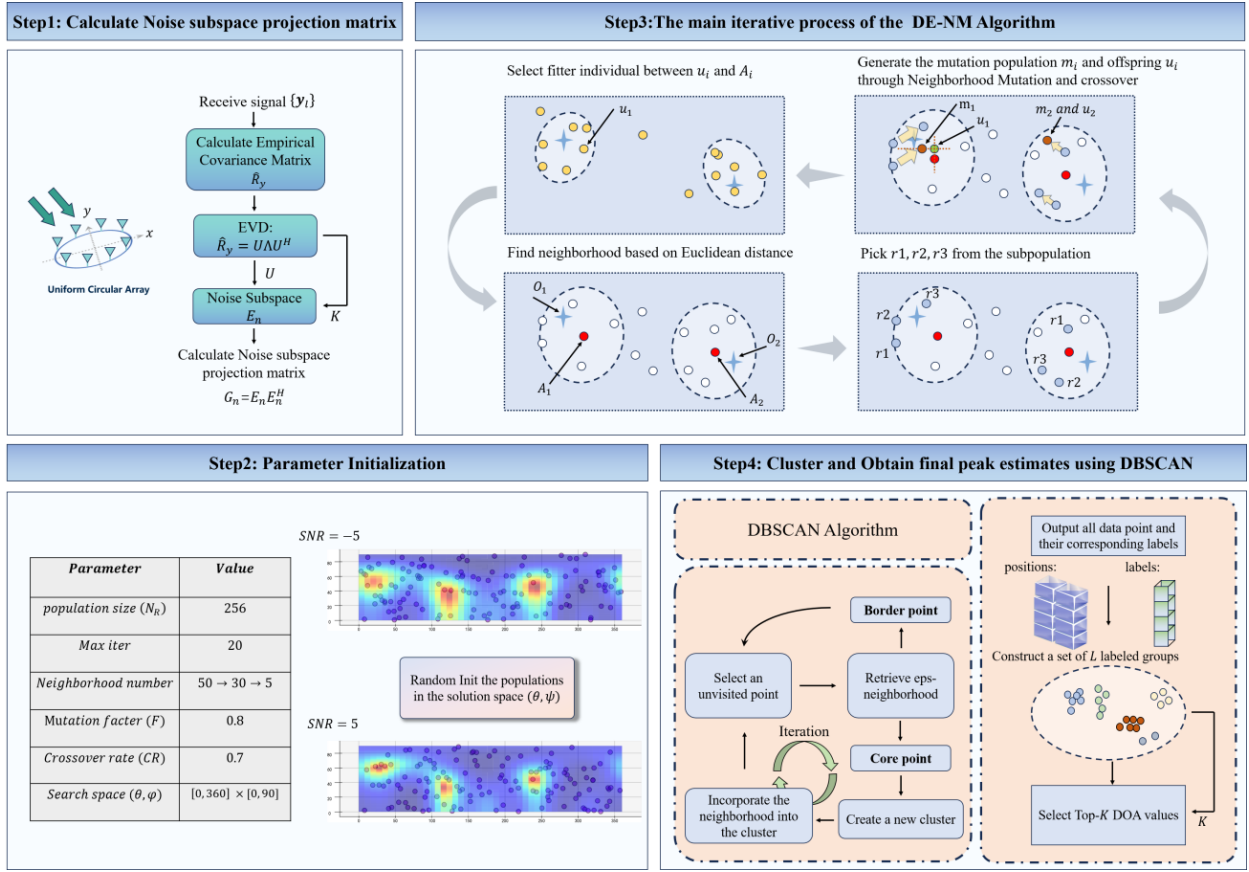

Figure 1. Framework of Differential Evolution Algorithm with Neighborhood Mutation.

Table 1 DE Algorithm - Pseudocode

**Algorithm 1** Basic Differential Evolution (DE) Algorithm
1: **Step 1:** Randomly initialize the population $\{x_i\}_{i=1}^P$ within the search space.
2: **Step 2:** Evaluate the fitness $f(x_i)$ for all individuals in the population.
3: **Step 3:** For iterations $iter = 1$ to $Max\_iter$ do
4:     **Step 3.1:** For each individual $x_i$, select distinct $x_{r1}, x_{r2}, x_{r3} \neq x_i$.
5:     **Step 3.2:** Mutation: $v_i = x_{r1} + F \cdot (x_{r2} - x_{r3})$
6:     **Step 3.3:** Crossover: generate $u_i$ from $x_i$ and $v_i$ using crossover rate $CR$.
7:     **Step 3.4:** Selection: if $f(u_i) \leq f(x_i)$, then $x_i \leftarrow u_i$
8: **Step 4:** Return the best individual $x_{best}$

Table 2 NDE Algorithm - Pseudocode

**Algorithm 2** Neighborhood-based Differential Evolution (NDE)
1: **Initialize:** Randomly initialize a population $\{x_i\}_{i=1}^P$ in the search space.
2: Evaluate the fitness $f(x_i)$ for all individuals.
3: **for** $iter = 1$ to $Max\_iter$ **do**
4:     **for** each individual $x_i$ **do**
5:         Find $m$ nearest neighbors of $x_i$ based on Euclidean distance to form $subpop_i$.
6:         Select $r_1, r_2, r_3 \in subpop_i$, with $r_1 \neq r_2 \neq r_3 \neq i$.
7:         **Mutation:** $\mathbf{v}_i = \mathbf{x}_{r_1} + F \cdot (\mathbf{x}_{r_2} - \mathbf{x}_{r_3})$
8:         **Crossover:** Generate trial vector $\mathbf{u}_i$ from $\mathbf{x}_i$ and $\mathbf{v}_i$ using rate $CR$.
9:         **Selection:** If $f(\mathbf{u}_i) \leq f(\mathbf{x}_i)$, then update $x_i \leftarrow \mathbf{u}_i$
10:     **end for**
11: **end for**
12: **return** the final population $\{x_i\}_{i=1}^P$

Table 3. Relative and Absolute Computational Complexity of 2D-MUSIC vs. Population-Based Algorithm While $N_R = 256$ and $Max\_iter = 20$

| MUSIC/Population-based Algorithm (MFLOPs) | $M = 12$ | $M = 32$ | $M = 128$ |
|---|---|---|---|
| $L = 1$ | 4.7/**2.0** (1:0.43) | 33.6/**6.5** (1:0.19) | 538.2/**85.2** (1:0.16) |
| $L = 3$ | 3.8/**1.9** (1:0.49) | 31.4/**6.2** (1:0.20) | 529.8/**83.9** (1:0.16) |

| $L = 10$ | **0.9**/1.4 (1:1.68) | 23.8/**5.0** (1:0.21) | 500.2/**79.4** (1:0.16) |

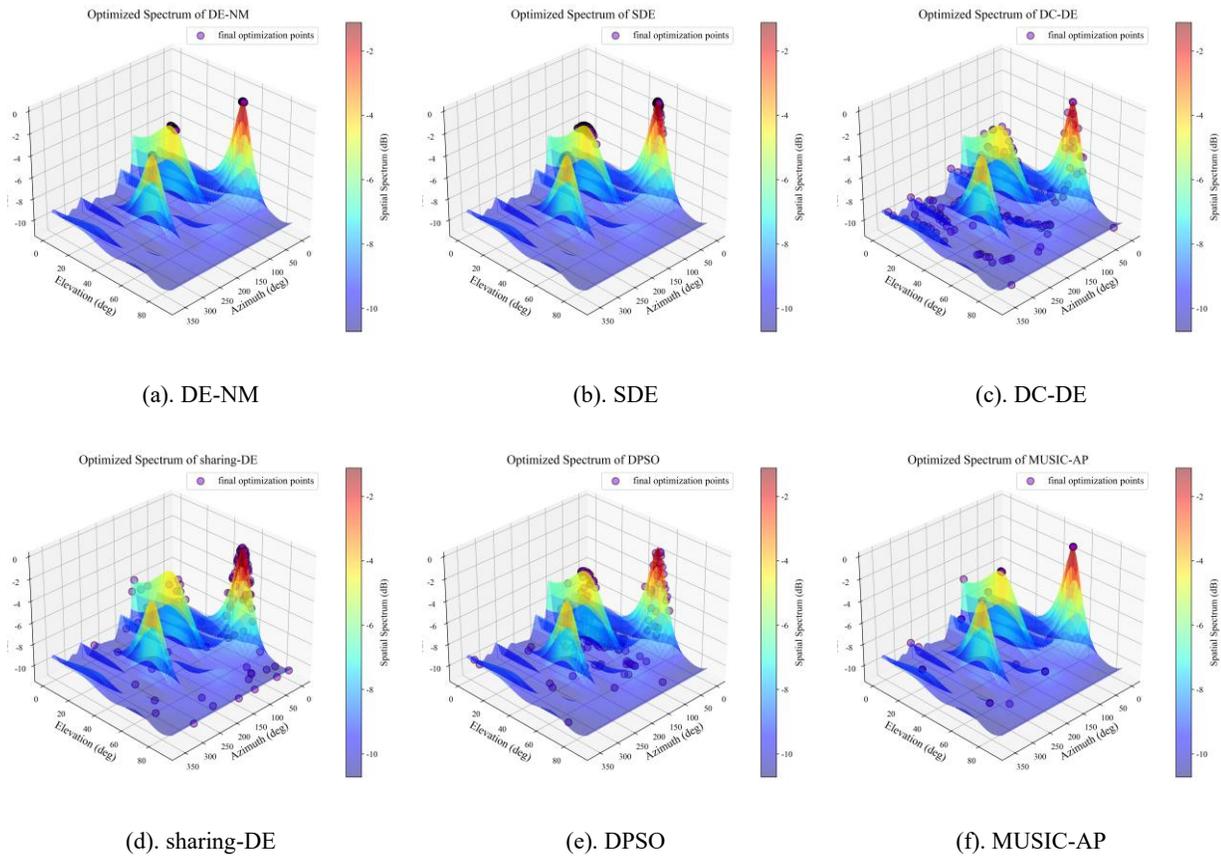

(a). DE-NM      (b). SDE      (c). DC-DE

(d). sharing-DE      (e). DPSO      (f). MUSIC-AP

Figure 2. The optimization results of different algorithms under the condition of SNR = –5 dB and snap=100.

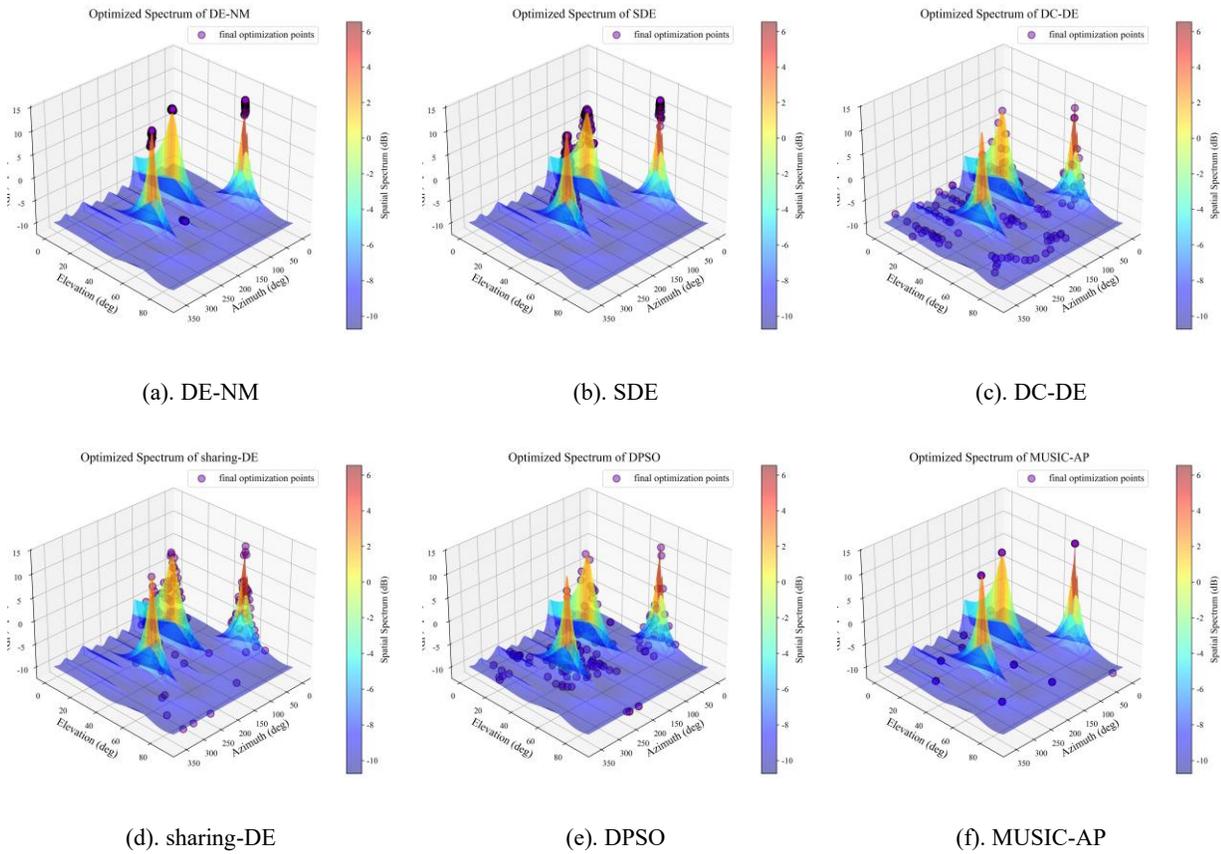

(a). DE-NM           (b). SDE           (c). DC-DE

(d). sharing-DE           (e). DPSO           (f). MUSIC-AP

Figure 3. The optimization results of different algorithms under the condition of SNR = 5 dB and snap=100.

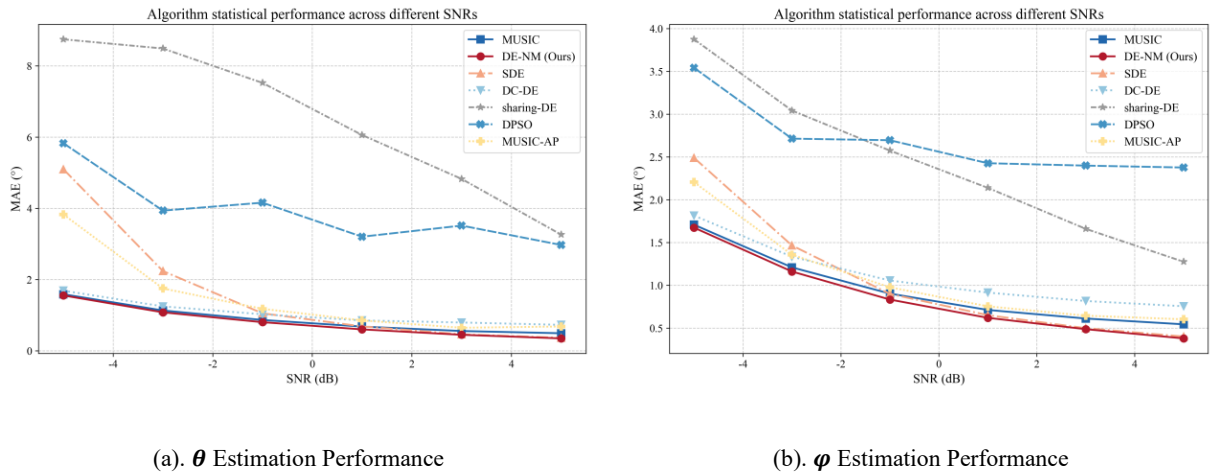

(a). $\theta$ Estimation Performance           (b). $\varphi$ Estimation Performance

Figure 4. The MAE performance of (a). $\theta$ and (b). $\varphi$ for various algorithms across different SNR levels.

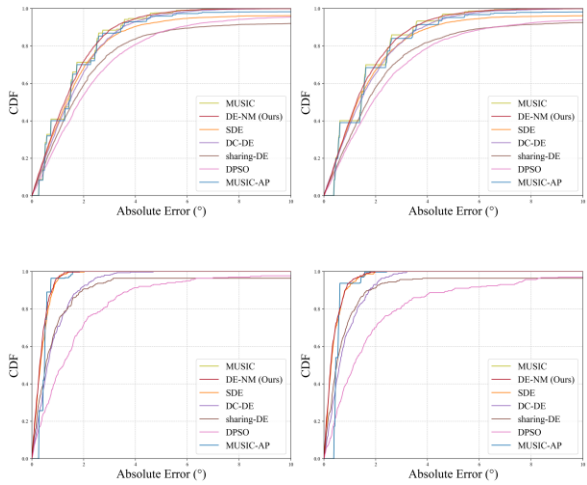

Figure 5. CDF of the absolute estimation errors for azimuth $\theta$ and elevation $\varphi$ based on 1,000 independent Monte Carlo trials.

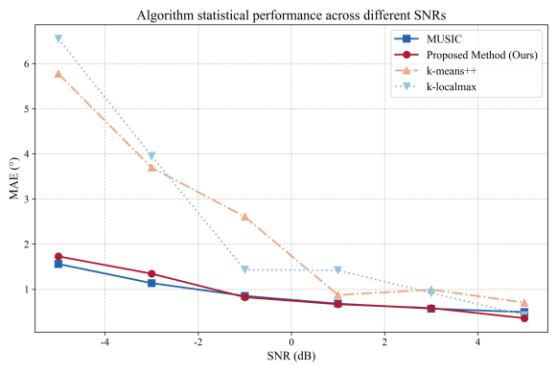

(a). MAE Performance for $\theta$

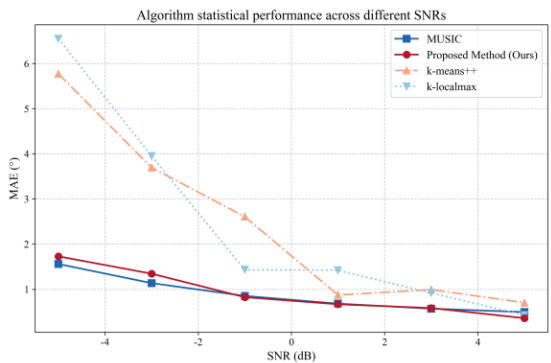

(b). MAE Performance for $\varphi$

Figure 6. MAE versus population size for azimuth $\theta$ and elevation $\varphi$ angles.

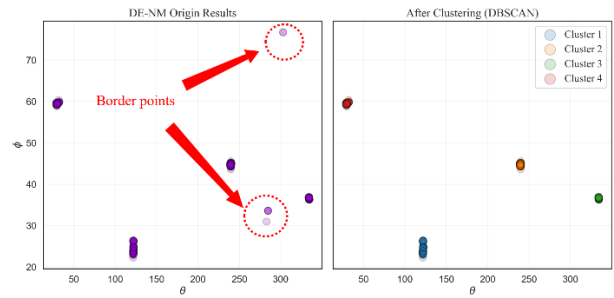

Figure 7. Population Distribution Results Before and After DBSCAN Clustering

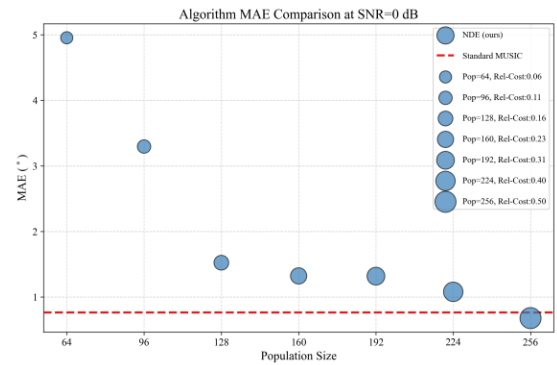

(a). MAE Performance for $\theta$

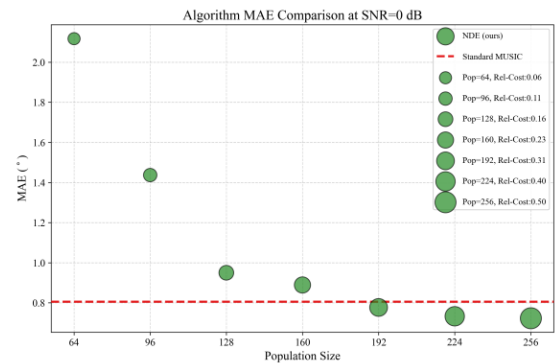

(b). MAE Performance for $\varphi$